%% file: Paper_epj.tex
\documentclass[epj]{svjour}
\usepackage[utf8]{inputenc}
\usepackage{lmodern}
\usepackage[T1]{fontenc}
\usepackage[english]{babel}
\usepackage{graphicx}
\usepackage{cite}
\usepackage{subfigure}
\usepackage{amsmath}
\usepackage{amssymb}
\newcommand{\ket}[1]{|#1\rangle}                           % Ket
\newcommand{\bra}[1]{\langle #1|}                          % Bra
\newcommand{\op}[1]{{\hat #1}}                             % Operator
\newcommand{\ov}[2]{\langle #1|#2\rangle}                  % Overlap
\newcommand{\brkt}[3]{\bra{#1} #2 \ket{#3}}                % Matrix element
\newcommand{\cc}[1]{\hat{c}^\dagger_{#1}}                  % Creation operator c
\newcommand{\ca}[1]{\hat{c}^{\,}_{#1}}                     % Annihilation operator
\newcommand{\qmoy}[1]{\bigl\langle #1 \bigr\rangle}        % Quantum expectation value <>
\newcommand{\ic}{\mathrm{i}}                               % Imaginary unit
\newcommand{\e}{\mathrm{e}}                                % e^1
\newcommand{\modul}[1]{\left\vert #1\right\vert}           % Module
\renewcommand{\d}{\mathrm{d}}                              % d for differential
\renewcommand{\Re}{\mathrm{Re}}                            % Real part
\newcommand{\moyE}{\mathbb{E}}                             % Statistical average E
\newcommand{\moy}[1]{\mathbb{E}\!\left[ #1\right]}         % Statistical average E[.]
\renewcommand{\H}{\hat{H}}                                     % Hamitonian H
\newcommand{\T}{\hat{T}}                                       % One-body part of H
\newcommand{\A}{\hat{A}}                                       % Generic operator A
\newcommand{\proptauinf}{\!\underset{\t\to+\infty}{\propto}\!} % Proportionnal when tau->+infinity
\newcommand{\eqtauinf}{\!\underset{\t\to+\infty}{=}\!}         % Equal when tau->+infinity
\newcommand{\Os}{\hat{O}_s}                                % One-body operators of the 2-body part of H
\newcommand{\ws}{\omega_s}                                 % Associated coefficients
\newcommand{\wt}[1]{\widetilde{#1}}                        % Widetilde
\newcommand{\ex}{\mathrm{ex}}                              % Excat
\newcommand{\PG}{\Psi_\ex}                                 % Exact eigenstate
\newcommand{\EG}{E_\ex}                                    % Exact eigenenergy
\newcommand{\tPG}{\wt{\Psi}_\ex}                           % Approximate eigenstate
\newcommand{\tP}{\wt{\Pi}}                                 % Biased weights
\newcommand{\R}{\rho}                                      % Single-particle transition density
\newcommand{\tEG}{\wt{E}_\ex}                              % Corresponding energy
\renewcommand{\P}{\hat{\mathcal{P}}}                       % Projector onto subspace
\newcommand{\Qnu}{\hat{\mathcal{Q}}^\nu}                   % Projector onto orthogonal subspace
\renewcommand{\t}{\tau}                                    % Imaginary time, tau
\newcommand{\Dt}{\Delta\tau}                               % Imaginary time step, Deltatau
\newcommand{\Om}{\Omega}                                   % Euler's angles, Omega
\renewcommand{\a}{\alpha}                                  % alpha
\renewcommand{\b}{\beta}                                   % beta
\begin{document}
\title{Constrained-Path Quantum Monte-Carlo Approach for Non-Yrast States Within the Shell Model} 
%Towards a Complete Spectroscopy of Nuclei with the Constrained-Path Quantum Monte-Carlo Approach for the Shell Model}
\author{J.~Bonnard\inst{1,2} \and O.~Juillet\inst{2}}
\institute{INFN, Sezione di Padova, I-35131 Padova, Italy \and LPC Caen, ENSICAEN, Universit{\'e} de Caen, CNRS/IN2P3, Caen, France}
\abstract{
The present paper intends to present an extension of the constrained-path quantum Monte-Carlo approach allowing to reconstruct non-yrast states in order to reach the complete spectroscopy of nuclei within the interacting shell model.
As in the yrast case studied in a previous work, the formalism involves a variational symmetry-restored wave function assuming two central roles.
First, it guides the underlying Brownian motion to improve the efficiency of the sampling.
Second, it constrains the stochastic paths according to the phaseless approximation to control sign or phase problems that usually plague fermionic QMC simulations.
Proof-of-principle results in the $sd$ valence space are reported.
They prove the ability of the scheme to offer remarkably accurate binding energies for both even- and odd-mass nuclei irrespective of the considered interaction.
 %Properties of nuclei; nuclear energy levels
}
\date{\today}
\titlerunning{Towards a Complete Spectroscopy of Nuclei with the CPQMC Approach}
\authorrunning{J.~Bonnard and O.~Juillet}

\maketitle
%
%\tableofcontents
%
%===================================================================================================================================
%===================================================================================================================================
\section{Introduction}
%===================================================================================================================================
%===================================================================================================================================
%
Quantum Monte-Carlo (QMC) methods represent appealing techniques to overcome the exponential complexity of many-body calculations by offering a systematic alternative to the diagonalization of the Hamiltonian.
In zero-temperature QMC formalisms, the lowest eigenstate of the Hamiltonian is indeed reconstructed from a stochastic reformulation of the imaginary-time Schr{\"o}dinger equation that reduces the many-body problem to a set of numerically solvable fluctuating one-body problems.
Nevertheless, except in very particular cases, fermionic QMC samplings generally suffer from the emergence of stochastic realizations with negative or complex weights cancelling the contributions of other realizations.
This pathology, the so-called sign or phase problem, plagues the simulations to the point of making them ineffective by causing an exponential collapse of the signal-to-noise ratio with imaginary time or system size.
QMC calculations with negative or complex weights fall within the broad class of NP-hard problems, i.e. problems of exponential complexity, which implies that there exists probably no algorithm with polynomial-time complexity \cite{I_Troyer_NPHard}.
Consequently, approximations have generally to be applied to manage the sign/phase problems.

In contrast with configuration-interaction techniques, QMC approaches rely on continuous many-body bases, against which the large variety of existing fermionic schemes may be classified in two main families.
First, the Green's-function Monte-Carlo (GFMC) \cite{I_GFQMC1,I_GFQMC2,I_GFQMC3} and diffusion Monte-Carlo (DMC) \cite{I_DMC1,I_DMC2,I_DMC3} methods stochastically explore the many-body coordinate space under the effect of the kinetic term of the imaginary-time propagator that is reinterpreted as random translations.
Hence, these approaches are suitable for interactions being diagonal in this basis, such as the Argonne series of nucleon-nucleon potentials \cite{I_Argonne}.
In presence of sign or phase problems, the fixed-node \cite{I_fixed_node1,I_fixed_node2,I_fixed_node3} and fixed-phase \cite{I_Ortiz_Fixed_phase} approximations are respectively used, leading thus to an upper-bound estimate of the exact ground-state energy.
The former controls the stochastic paths by preventing the random walkers from crossing the nodal surface of a given trial wave function.
The latter holds the phase of the walkers equal to the phase of a complex trial state to ensure positive weights during the Brownian motion.
It reduces to the fixed-node constraint when real wave functions are considered.
GFMC and DMC methods nowadays provide powerful theoretical frameworks for accurate microscopic calculations in various areas of physics.
For instance, they have allowed pioneering \textit{ab initio} studies of the low-lying structure of light nuclei \cite{I_Pieper_GFMC} and neutron matter \cite{I_Carlson_GFMC} from two- and three-body nuclear forces, including investigations of the Hoyle state in ${}^{12}$C that is of crucial importance for nucleosynthesis of carbon in stars \cite{I_Lusk_Hoyle}.
Furthermore, the auxiliary-field diffusion Monte-Carlo method \cite{I_AFDMC1,I_AFDMC2}, that generalizes DMC by incorporating a stochastic treatment of the spin-isospin degrees of freedom, has recently been used to carry out the first QMC simulations with modern realistic interactions derived from chiral effective field theory \cite{I_QMCChiEFT}.

In this work, we mainly focus on the second family of QMC approaches in which the random walk takes place in product-state spaces.
These formalisms may be applied to many-body problems defined from any finite and discrete single-particle basis. 
The underlying strategy is to replace the interacting part of the propagator by stochastic one-body fields, commonly \textit{via} the Hubbard-Stratonovich transformation \cite{I_Hubb_Strato_transf1,I_Hubb_Strato_transf2}, in such a way that the imaginary-time dynamics is exactly reproduced in average.
The auxiliary-field quantum Monte-Carlo (AFQMC) approach \cite{I_AFQMC_1,I_AFQMC_2} is the most known QMC method of this family and has also been used in a wide range of applications.
An efficient way for addressing the sign problem is furnished by the constrained-path Monte-Carlo approach \cite{I_Zhang_CPMC,I_Zhang_CPMC_book}. 
Indeed, it incorporates in standard AFQMC samplings an approximation whose principle is similar to the fixed-node one but operating within the Slater determinant manifold instead of the coordinate space.
In addition, the phaseless QMC scheme --- detailed in this paper --- has been specially proposed in quantum chemistry to handle the phase problem \cite{I_Zhang_Phaseless1,I_Zhang_Phaseless2,I_Zhang_Phaseless3}.

The interacting nuclear shell model \cite{I_Caurier_SM} constitutes a typical example of calculations whose exponential complexity may be circumvented thanks to a stochastic reformulation.
In this picture, the valence protons and neutrons beyond an inert magic core are confined in one or more active shells and interact through an effective two-body residual potential.
Traditionally, the Schr{\"o}dinger equation is solved by diagonalization of the Hamiltonian matrix in the set of all the accessible configurations, which faces the prohibitive scaling of the many-body basis dimension with the size of the single-particle space and nucleon number.
The shell model Monte-Carlo (SMMC) method \cite{I_Koonin_SMMC} that remains to date the principal AFQMC approach of the shell model, has thus enabled to investigate systems out of reach of conventional shell model calculations without severe truncations of the configuration space.
The studies of $pf$-shell nuclei \cite{I_SMMC_pfGT}, rare-earth nuclei \cite{I_SMMC_rare}, heavy-deformed nuclei \cite{I_SMMC_defnuc}, Gamow-Teller distributions \cite{I_SMMC_pfGT}, temperature dependence of pairing correlations \cite{I_SMMC_Tpair}, double-$\b$ decay \cite{I_SMMC_bb}, and microscopic computations of level densities \cite{I_SMMC_leveldens1,I_SMMC_leveldens2,I_SMMC_leveldens3,I_SMMC_leveldens4}, are concrete applications of the SMMC method.
With schematic effective interactions, e.g. pairing-plus-quadrupole potentials, this approach yields exactly the properties of even-even and $N=Z$ odd-odd nuclei at zero and finite temperature \cite{I_SMMC_PhP}.
However, since it is based on a standard AFQMC sampling, the SMMC method inevitably undergoes phase problems with realistic effective interactions or for other kind of nuclei.
To avoid this pathology and guarantee convergent simulations, a class of modified Hamiltonians where the interactions causing the problem are artificially reduced is defined.
The physical observables are then extracted by extrapolating these non-physical SMMC results to recover the original Hamiltonian \cite{I_SMMC_PhExt}.

Besides the phase problem, another limitation of the SMMC approach lies in its incapacity to achieve a detailed spectroscopy of nuclei, albeit informations on excited states can be deduced from response functions.
A first attempt to describe the yrast spectroscopy in QMC treatments of the shell model was suggested by Puddu \cite{I_Puddu_SpectroQMC1,I_Puddu_SpectroQMC2,I_Puddu_SpectroQMC3,I_Puddu_SpectroQMC4}.
As the SMMC method, this approach corresponds to the AFQMC scheme, but with walkers partially or fully projected onto the relevant quantum numbers.
Therefore, it is also plagued by phase problems and only schematic residual interactions have been considered.
On the other hand, low-lying states can be reconstructed by means of the so-called Monte-Carlo shell model technique \cite{I_MCSM_I1,I_MCSM_I2} that does not utilize the AFQMC stochastic process to sample eigenstates, but to generate a subspace into which the Hamiltonian matrix is diagonalized.
Given that such calculations are not subject to explicit manifestations of sign or phase problems, they may be performed in very large configuration spaces to obtain variational estimates of nuclear properties \cite{I_MCSM_II}.
This method has been used to study the shape coexistence in ${}^{56}$Ni \cite{I_MCSM_56Ni}, the structure of neutron-rich exotic nuclei \cite{I_MCSM_exotic}, for example, and has been recently applied within an \textit{ab initio} context to no-core calculations for the beryllium isotopes \cite{I_MCSM_Nocore}.

In a previous work \cite{PRL}, we have reported a new QMC scheme for the shell model providing nearly exact yrast spectroscopy with a well-controlled phase problem for both even- and odd-mass nuclei and for any interaction.
The formalism, derived from the phaseless QMC scheme, relies on a symmetry-restored trial wave function to guide as well as to constrain the Brownian motion of Slater determinants.
The present paper pursues the following objective: 
To reach the full low-lying spectroscopy of nuclei through an extension of the approach to non-yrast states.
Sect. \ref{Sect_2} reviews the phaseless QMC formalism discussing in details the origin of the phase problem and the constraint to manage it. Sect. \ref{Sect_3} displays the extension to the excited states and proof-of-principle results for $sd$-shell nuclei are reported in Sect. \ref{Sect_4}.
This work is finally summarized in Sect. \ref{Sect_5}.
%
%
%
%
%===================================================================================================================================
%===================================================================================================================================
\section{General features of the phaseless QMC scheme}\label{Sect_2}
%===================================================================================================================================
%===================================================================================================================================
%
Implementing an AFQMC-like approach requires the two-body Hamiltonian $\H$ to be cast in a quadratic form of one-body operators $\T$ and $\{\Os\}$:
\begin{subequations}\label{QuadFormH}
\begin{equation}
  \H =\T -\sum_s \ws\Os^2,
\end{equation}
\begin{equation}
   \T =\sum_{i,j} T_{ij} \cc{i}\ca{j},\quad \Os =\sum_{i,j} (O_s)_{ij} \cc{i}\ca{j},
\end{equation}
\end{subequations}
where $\cc{i}$ ($\ca{i}$) is the creation (annihilation) operator of a fermion in the single-particle state $\ket{i}$ of a finite-size discrete orthonormal basis.
A possible general procedure to rewrite any Hamiltonian may be found in Ref. \cite{I_Zhang_CPMC_book}.
As discussed in Ref. \cite{I_SMMC_PhP} in the case of the nuclear shell model, there is substantial freedom in doing so.
Here, we adopt the isospin density decomposition of $\H$ given by Ref. \cite{II_Lang_noPPSMMC2} that exhibits a remarkable property:
The $\T$ and $\{\Os\}$ operators do not mix neutrons and protons, which interestingly impacts the speed of QMC simulations.

Zero-temperature QMC methods rely on the imaginary-time evolution to project an initial wave function $\ket{\Phi_0}$ onto the lowest-energy eigenstate $\ket{\PG}$ of the Hamiltonian such as $\ov{\Phi_0}{\PG}\neq0$:
\begin{equation} \label{it_Prop}
  \lim_{\t\to +\infty} \e^{-\t\H} \ket{\Phi_0} \propto \ket{\PG} .
\end{equation}
The second-order Trotter-Suzuki breakup \cite{II_TS1,II_TS2} and the so-called Hubbard-Stratonovich transformation \cite{I_Hubb_Strato_transf1,I_Hubb_Strato_transf2} then allow to reformulate stochastically the propagator for a short imaginary-time step $\Dt$
\begin{equation} \label{Int_AFQMC}
  \e^{-\Dt\H} = \int\!\! \d\vec{\eta} \, p(\vec{\eta})\,\hat{\mathcal{U}}(\vec{\eta}),
\end{equation}
where $\vec{\eta}$ is a vector of random variables $\{\eta_s\}$, the auxiliary fields, distributed according to a normal Gaussian distribution $p$.
As exponential of one-body operators, the stochastic propagator
\begin{equation} \label{AFQMCdyn}
  \hat{\mathcal{U}} (\vec{\eta})=\e^{-\frac{\Dt}{2}\T}\e^{\sum_s\eta_s\sqrt{2\ws\Dt}\Os}\e^{-\frac{\Dt}{2}\T} ,
\end{equation}
transforms a Slater determinant $\ket{\Phi_\t}$ at $\t$ to a new one $\ket{\Phi_{\t+\Dt}}=\hat{\mathcal{U}}\ket{\Phi_\t}$ involved in the sampling of the exact state at $\t+\Dt$.
Accordingly, the wave function resulting from the exact evolution during $\Dt$ of $\ket{\Phi_\t}$ is reinterpreted as the coherent statistical average $\moy{\cdot}$ of independent-fermion states $\ket{\Phi_{\t+\Dt}}$, i.e. $\exp(-\Dt\H) \ket{\Phi_\t} = \moy{\ket{\Phi_{\t+\Dt}}}$, and finally the eigenstate of $\H$ is recovered as
\begin{equation}
  \ket{\PG} \proptauinf \moy{\ket{\Phi_\t}}.
\end{equation}

As mentioned previously, the phase problem embodies the main difficulty of QMC samplings.
Its origin lies in the possibility for the phase of the overlap $\ov{\PG}{\Phi_\t}$ to change.
Indeed, extracting $\ket{\PG}$ becomes strongly compromised as soon as the Brownian motion generates a large proportion of walkers corresponding to different phases and inducing a mutual cancellation of overlaps, that is such as $\moy{\ov{\PG}{\Phi_\t}} \approx 0$.
To understand this, let us suppose that such a population is encountered at time $\t^*$. 
In the complex $\ov{\PG}{\Phi_\t}$ plane, they form a set of points whose center of mass exactly merges with the origin.
Obviously, such walkers degrade the signal-to-noise ratio: They do not contribute to the reconstruction of $\ket{\PG}$, but only to the statistical errors.
Their average $\ket{\Psi_\perp}$ is by definition a state orthogonal to $\ket{\PG}$, and for any time interval $\Dt$, $\ov{\PG}{\Psi_\perp} = 0$ implies
\begin{equation} \label{phase_const_PP}
  \begin{split}
    \brkt{\PG}{\e^{-\Dt\H}}{\Psi_\perp} &= \e^{-\Dt\EG} \ov{\PG}{\Psi_\perp} = 0 \\
    & = \moy{\ov{\PG}{\Phi_{\t^*+\Dt}}} ,
  \end{split}
\end{equation}
with $\EG$ the exact energy.
This means that the propagation during any time $\Dt$ of the pathological realizations at $\t^*$ yields walkers that also have collectively a zero mean overlap with $\ket{\PG}$.
Consequently, the number of realizations that do not contribute to the sampling increases with the imaginary time, whereas in parallel, the proportion of those that effectively participate exponentially decreases.
This phenomenon is not incompatible with a first moment equal to $\ket{\PG}$, but inexorably entails a divergence of both the mean error and the quadratic mean error.
The exponential behaviour of the decrease of the signal-to-noise ratio, as well as the fact that the number of walkers originating this decrease at any time $\t^*$ can be very small (even equal to one, see below), clearly attest to the extreme severity of the phase problem in QMC simulations.

A pathological population at $\t^*$ causing the phase problem may, for example, result from the stochastic propagation of a walker that has previously crossed at a time $\t'$ the nodal surface of the exact state:
\begin{equation}
  \begin{split}
    \brkt{\PG}{\e^{-(\t^*-\t')\H}}{\Phi_{\t'}} &= \e^{-(\t^*-\t')\EG} \ov{\PG}{\Phi_{\t'}} = 0 \\
    &= \bra{\PG}\moy{\ket{\Phi_{\t^*}}} = \ov{\PG}{\Psi_\perp}.
  \end{split}
\end{equation}
When a sampling with Slater determinants composed of real orbitals is possible, this scenario represents the only way for the phase of $\ov{\PG}{\Phi_\t}$ to vary during the random walk.
This is the one traditionally invoked to explain the origin of sign problems in position-space-based QMC approaches \cite{I_GFQMC1} and AFQMC schemes \cite{II_Zhang_SPPP}.
On the other hand, with complex single-particle wave functions, the phase change does not force the walkers to pass through the nodal surface.

Generally, managing the phase problem requires to resort to approximations, such as the constrained-path one \cite{I_Zhang_CPMC,I_Zhang_CPMC_book}, based on a trial wave function $\ket{\Psi_T}$ not orthogonal to the exact state.
In this case, $\ket{\PG}$ can be viewed as stemming from the long-time stochastic evolution of $\ket{\Psi_T}$, and so
\begin{equation} \label{AFQMC_PP}
  \ov{\PG}{\Phi}\proptauinf \brkt{\Psi_T}{\e^{-\t\H}}{\Phi} = \moy{\ov{\Psi_T}{\Phi_\t}} ,
\end{equation}
with $\ket{\Phi_0} = \ket{\Phi}$.
These approximations take advantage of the link \eqref{AFQMC_PP} between the overlap of the walkers with the exact state and the one with the trial wave function in order to discard the pathological realizations from the sampling.

The most intuitive approach to eliminate the problematic realizations would be to retain only the walkers that correspond to the same phase for their overlaps with $\ket{\Psi_T}$.
However, if the random walk tends to cover the whole complex $\ov{\Psi_T}{\Phi_\t}$ plane, nearly all the Slater determinants would be rejected from the sampling, which would tremendously degrade the distribution of the overlaps compared to the original one arising from the dynamics.
The principal issue is, hence, to find a good compromise between the no-constrained form of the distribution that ensures an exact reconstitution of the desired stationary state, and the necessity to control the phase problem.
Nevertheless, with Slater determinants built from real single-particle states, the constraint $\ov{\Psi_T}{\Phi_\t} > 0$ is acceptable.
This matches the so-called constrained-path approximation \cite{I_Zhang_CPMC,I_Zhang_CPMC_book} that reveals very efficient in investigating lattice models of strongly-correlated electrons, such as the Hubbard model.
A natural extension of this approximation for complex orbitals is to enable the phase of $\ov{\Psi_T}{\Phi_\t}$ to vary while avoiding pathological realizations to occur by maintaining the Brownian motion within one of the half-planes, for instance, $\Re \ov{\Psi_T}{\Phi_\t} > 0$, $\forall\t$.
The center-of-mass of any subset of points thus never coincides with the origin.
Concretely, this approximation is easily implemented by following the commonly used strategy of importance sampling.
The probability distribution according to which are generated the realizations is then modified to include in the sampling the overlap of the walkers with the trial wave function:
\begin{equation} \label{approx_compCP1}
  \moy{\ket{\Phi_\t}} = \moyE_\Pi \left[ \e^{\ic\theta_\t} \dfrac{\ket{\Phi_\t}}{\ov{\Psi_T}{\Phi_\t}} \right].
\end{equation}
$\theta_\t$ denotes the phase of $\ov{\Psi_T}{\Phi_\t}$ and $\moyE_\Pi$ the average with the weight $\Pi=|{\ov{\Psi_T}{\Phi_\t}}|$.
In the constrained-path approximations, all stochastic paths corresponding to $|\theta_\t|>\pi/2$ are discarded and the reconstruction scheme \eqref{approx_compCP1} is replaced by
\begin{equation} \label{approx_compCP}
  \moy{\ket{\Phi_\t}} \to \mathbb{E}_{\tP} \left[ \dfrac{\ket{\Phi_\t}}{\ov{\Psi_T}{\Phi_\t}} \right] ,
\end{equation}
where $\tP_\t = \max\bigl\lbrace 0\,;\Re\ov{\Psi_T}{\Phi_\t}\bigr\rbrace$.
\begin{figure}[!t]
\begin{center}
\includegraphics[scale=0.318]{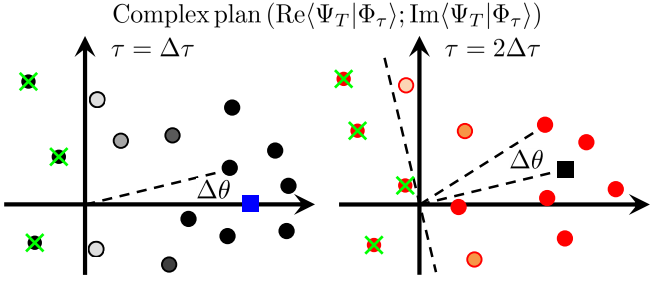}
\caption{\label{Fig_approx_Ph}\textbf{(Color online)} Schematic illustration of the principle of the phaseless approximation. At time $\Dt$ (left panel), the random walk generates from the initial Slater determinant $\ket{\Phi_0}$ (blue square) a collection of realizations (black bullets) whose weights, indicated by the transparency of the points, depend on the dephasing $\Delta\theta$ of the their overlaps with $\ket{\Psi_T}$. The realizations undergoing a phase change $\modul{\Delta\theta}>\pi/2$ are eliminated. At time $2\Dt$ (right panel), every previous point produces a new set of realizations (red bullets), as shown for one (black square). Applying the constraint iteratively, the centroid of the global collection obtained after a large imaginary time, associated with the approximate ground state $\ket{\tPG}$, necessarily belongs to the half-plane $\Re\ov{\Psi_T}{\Phi_\t}>0$, even if some realizations populate the other half-plane.}
\end{center}
\end{figure}
Within the framework of the standard AFQMC approach, the phaseless approximation \cite{I_Zhang_Phaseless1,I_Zhang_Phaseless2,I_Zhang_Phaseless3} relies on a similar underlying idea, but is far less restrictive.
As illustrated in Fig. \ref{Fig_approx_Ph}, it schematically aims at producing a collection of points in the complex $\ov{\Psi_T}{\Phi_\t}$ plane around the initial phase by adequately modifying their weights throughout the propagation in accordance with the phase change $\Delta\theta_\t$ between $\t$ and $\t+\Dt$.
Such a constraint is also easily implemented by means of an importance sampling with biased weights $\tP$ evolving as
\begin{subequations} \label{Ph_approx}
\begin{equation} \label{Ph_approx_PP}
  \tP_{\t+\Dt} = \tP_\t \modul{\dfrac{\ov{\Psi_T}{\Phi_{\t+\Dt}}}{\ov{\Psi_T}{\Phi_\t}}} \max\Bigl\lbrace 0\,;\cos\Delta\theta_\t\Bigr\rbrace ,
\end{equation}
the dephasing $\Delta\theta_\t$ being defined from the ratio of overlaps at $\t$ and $\t+\Dt$
\begin{equation} \label{Ph_approx_dtheta_PP}
  \Delta\theta_\t = \theta_{\t+\Dt}-\theta_\t =\mathrm{arg} \dfrac{\ov{\Psi_T}{\Phi_{\t+\Dt}}}{\ov{\Psi_T}{\Phi_\t}} .
\end{equation}
\end{subequations}
Thanks to the factor $\cos(\Delta\theta_\t)$, the more the phase of the overlap with the trial state varies during $\Dt$, the more the weight of the associated walker is reduced, up to be put to zero when $\modul{\Delta\theta_\t}$ becomes larger than $\pi/2$.
Iterating the process, some points reach the region $\Re\ov{\Psi_T}{\Phi_\t}<0$ --- the probability distribution is therefore less degraded than with the approximations described previously ---, but the centroid persists in the other half-plane because their weights gradually decrease over the imaginary time.
In this respect, the constraint \eqref{Ph_approx} offers a good compromise between the need to control the phase problem and the conservation of the initial form for the distribution.
Schematically, in the many-body Hilbert space, a collection of walkers is thus generated in the vicinity of the trial wave function and represents a free of phase problem sampling of an approximate eigenstate of the Hamiltonian.

The so-called phaseless QMC scheme \cite{I_Zhang_Phaseless1,I_Zhang_Phaseless2,I_Zhang_Phaseless3} suggested by Zhang and Krakauer in quantum chemistry is based on the approximation \eqref{Ph_approx}.
Moreover, the efficiency of the AFQMC method is improved by including directly within the random walk an importance sampling with the overlap between the walkers and a trial wave function $\ket{\Psi_T}$ as importance function.
The exact state is first reformulated as a weighted average of Slater determinants,
\begin{equation} \label{ansatz_Ph}
  \ket{\PG}\proptauinf  \moy{\Pi_\t\dfrac{\ket{\Phi_\t}}{\ov{\Psi_T}{\Phi_\t}}}.
\end{equation}
Here, the factor $\Pi$ is introduced to ensure that the stochastic scheme is equivalent to the exact imaginary-time propagation.
The equations of motion may be obtained through the standard Hubbard-Stratonovich procedure depicted above as long as the $\Os$ operators in Eq. \eqref{QuadFormH} are shifted by their local estimates $\qmoy{\Os}_{\Psi_T\Phi_\t} = \brkt{\Psi_T}{\Os}{\Phi_\t} / \ov{\Psi_T}{\Phi_\t}$.
This finally leads to an exponential increment for the weight \cite{I_Zhang_Phaseless1,Proc}
\begin{equation} 
  \Pi_{\t+\Dt} = \Pi_\t \exp\Bigl(-\Dt\qmoy{\H}_{\Psi_T\Phi_\t}\Bigr),
\end{equation}
and to the following stochastic propagator:
\begin{subequations} \label{ev_disc_orb}
\begin{equation} 
  \hat{\mathcal{U}}_T(\vec{\eta}) = \e^{-\frac{\Dt}{2}\T} \e^{\sum_s\Delta h_{T,s}\Os} \e^{-\frac{\Dt}{2}\T},
\end{equation}
with
\begin{equation}
  \Delta h_{T,s}=2\Dt\ws \qmoy{\Os}_{\Psi_T\Phi_\t} +\eta_s \sqrt{2\ws\Dt}.
\end{equation}
\end{subequations}
Compared to the standard AFQMC scheme, see Eq. \eqref{AFQMCdyn}, the trial state $\ket{\Psi_T}$ now takes part in the dynamics \textit{via} the local estimates of the $\op{O}_s$, and thereby steers the random walk towards a region of the Slater determinant manifold where their importance in the sampling of $\ket{\PG}$ is expected to be large.
Note that since when $\Dt\to\d\t$ the quantities $\eta_s\sqrt{\Dt}$ become infinitesimal increments $\d W_s$ of Wiener processes in the Ito formalism of stochastic calculus \cite{II_Gardiner_StochCalc}, one may check that the one-body representation of $\hat{\mathcal{U}}_T$ governing the Brownian dynamics of the occupied orbitals transforms into the stochastic differential equations (3) of Ref. \cite{PRL}.
Finally, adapting the constraint \eqref{Ph_approx}, an approximate eigenstate $\ket{\tPG}$ of the Hamiltonian is reconstructed as
\begin{subequations} \label{Ph_scheme}
\begin{equation} \label{Ph_scheme_ket}
  \ket{\tPG} \proptauinf \moy{\tP_\t\dfrac{\ket{\Phi_\t}}{\ov{\Psi_T}{\Phi_\t}}},
\end{equation} %\widetilde{\e^{-\t\H}} \ket{\Phi_0} = 
where the biased weights vary with the imaginary time according to
\begin{equation} \label{Ph_scheme_Pi}
  \tP_{\t+\Dt} = \tP_\t \exp\Bigl(-\Dt\Re\qmoy{\H}_{\Psi_T\Phi_\t} \Bigr) \max\bigl\lbrace 0\,;\cos\Delta\theta_\t\bigr\rbrace.
\end{equation}
\end{subequations}
%
%the dephasing $\Delta\theta_\t$ being given by Eq. \eqref{Ph_approx_dtheta_PP}.
In conclusion, let us emphasize that no assumption has been made in what precedes regarding the form of $\ket{\Psi_T}$ --- except $\ov{\Psi_T}{\PG} \neq 0$ ---, so that it can be a correlated wave function.
%
%
%
%===================================================================================================================================
%===================================================================================================================================
\section{Extension of the phaseless QMC approach for non-yrast states}\label{Sect_3}
%===================================================================================================================================
%===================================================================================================================================
%
%
In the present section, we consider the reconstruction of the $\nu$-th excited state $\ket{\PG^{JM\nu}}$ in the angular-momentum channel $J, M$.
%
%
%=======================================================================================================
\subsection{Adaptation of the QMC scheme}
%=======================================================================================================
%
Adapting the phaseless QMC method to the sampling of non-yrast states requires the initial and trial wave functions to satisfy particular properties, besides the fact that both have to share the angular-momentum quantum numbers of $\ket{\PG^{JM\nu}}$.
First, the imaginary-time propagation has to be initialized by a state $\ket{\Psi^{JM\nu}_0}$ whose overlap with the desired eigenstate is not equal to zero, i.e. $\ov{\PG^{JM\nu}}{\Psi^{JM\nu}_0}$ $\neq 0$.
This condition may be fulfilled by symmetry restoration from a general Slater determinant $\ket{\Phi^\nu_0}$:
\begin{equation} \label{Ansatz_SEMF_n}
  \ket{\Psi^{JM\nu}_0}= \sum_{K=-J}^J C^{J\nu}_K\,\op{P}^J_{MK} \ket{\Phi^\nu_0}.
\end{equation}
The projection operators $\op{P}^J_{MK}$ are weighted averages of the Euler's angle $\Om$ parametrization of rotations $\op{U}_\Om$ \cite{III_Villar}:
\begin{equation} \label{Proj_PJKM}
  \op{P}^J_{MK} = \dfrac{2J+1}{8\pi^2} \int\!\! \d\Om\,D^{J*}_{MK}(\Om)\op{U}_\Om ,
\end{equation}
$D^{J}_{MK}$ denoting Wigner's $D$-functions.
The linear combination mixing with amplitudes $\{C^{J}_K\}$ all the possible values of the spin projection $K$ in the intrinsic frame is introduced to guarantee vectors \eqref{Ansatz_SEMF_n} to constitute a standard basis for the total angular momentum.
Furthermore, in order to ensure convergence to the wanted state, the trial state $\ket{\Psi^{JM\nu}_T}$ has to obey $\ov{\PG^{JM\nu}}{\Psi^{JM\nu}_T}$ $\neq 0$, but also to be strictly orthogonal to all the eigenstates of lower energies, that is $\ov{\PG^{JM\a}}{\Psi^{JM\nu}_T}$ $= 0$ for all $\a=0$ (yrast state) to $\nu-1$.
It is possible to define such a wave function as
\begin{equation} \label{Ansatz_SEMFExc_n0}
  \ket{\Psi_T^{JM\nu}} = (\hat{1}-\P^{JM\nu}_\ex) \ket{\Psi^{JM\nu}_0},
\end{equation}
by introducing the projector $\P^{JM\nu}_\ex = \sum_{\a=0}^{\nu-1}\ket{\PG^{JM\a}}\bra{\PG^{JM\a}}$ onto the subspace spanned by the eigenstates of lower energies.

From the ansatz \eqref{Ansatz_SEMF_n} and \eqref{Ansatz_SEMFExc_n0}, the exact energy $\EG^{J\nu}$ of the $\nu$-th excited state $\ket{\PG^{JM\nu}}$ of spin $J$ may in principle be obtained after a large enough imaginary time as
\begin{equation}\label{Ex_nonyrast_Eg}
  \EG^{J\nu}\eqtauinf \dfrac{\brkt{\Psi_T^{JM\nu}}{\H\e^{-\t\H}}{\Psi^{JM\nu}_0}}{\brkt{\Psi_T^{JM\nu}}{\e^{-\t\H}}{\Psi^{JM\nu}_0}} 
  = \dfrac{\brkt{\Psi_T^{J\nu}}{\H\e^{-\t\H}}{\Phi^\nu_0}}{\brkt{\Psi_T^{J\nu}}{\e^{-\t\H}}{\Phi^\nu_0}}.
\end{equation}
We have written 
\begin{equation} \label{Trial_stateExc0}
  \ket{\Psi_T^{J\nu}} = \sum_{K} C^{J\nu*}_K  \ket{\Psi_T^{JK\nu}},
\end{equation}
the new state that emerges from rotational invariance and from the obvious relation
\begin{equation} \label{id_exproj}
  (\hat{1}-\P^{JM\nu}_\ex)\op{P}^J_{MK} = \op{P}^J_{MK}(\hat{1}-\P^{JK\nu}_\ex).
\end{equation}
Note that $\ket{\Psi_T^{J\nu}}$ being a linear combination of vectors \eqref{Ansatz_SEMFExc_n0} projected onto different quantum numbers $K$, it remains orthogonal to the lower eigenstates of the Hamiltonian. 
Numerically, computing Eq. \eqref{Ex_nonyrast_Eg} \textit{via} an \emph{exact} stochastic sampling actually risks to converge to an indeterminate form `$0/0$'.
Indeed, in contrast to the trial wave function, the initial Slater determinant $\ket{\Phi^\nu_0}$ possibly has a non-zero overlap with lower eigenstates.
In such a case, its propagation inescapably yields the lowest one that is perpendicular to $\ket{\PG^{JM\nu}}$.
However, for realistic shell model effective interactions, an exact imaginary-time evolution is utopian.
Within the phaseless QMC scheme, the convergence is guaranteed by the control of the phase problem forcing the population of walkers to have collectively a real and positive overlap with $\ket{\Psi_T^{J\nu}}$.
Nevertheless, the unreachable projector $\P^{JM\nu}_\ex$ embedded in the wave function \eqref{Ansatz_SEMFExc_n0} has to be approximated to obtain an operational approach.
In the following, we impose the orthogonality of $\ket{\Psi_T^{J\nu}}$ with the trial states $\ket{\Psi_T^{JM\a}}$ ($\a=0\to\nu-1$) adopted for the reconstruction of the lower-energy levels in the $J$ sector:
\begin{subequations} \label{Ansatz_SEMFExc_n}
\begin{equation} 
  \ket{\Psi_T^{JM\nu}} = (\hat{1}-\P^{JM\nu}_T) \ket{\Psi^{JM\nu}_0}.
\end{equation}
Here, $\P^{JM\nu}_T$ denotes the projector onto the subspace spanned by the states $\ket{\Psi_T^{JM\a}}$, or equivalently the non-orthonormalized vectors $\{\ket{\Psi_0^{JM\a}}\}$.
Thus, the projector $\P^{JM\nu}_T$ can be written as
\begin{equation} \label{proj_T}
  \P^{JM\nu}_T = \sum_{\a,\b=0}^{\nu-1} \ket{\Psi^{JM\a}_0} (F^{-1})_{\a\b} \bra{\Psi^{JM\b}_0},
\end{equation}
\end{subequations}
where $F$ stands for the overlap matrix $F_{\a\b}=\ov{\Psi^{JM\a}_0}{\Psi^{JM\b}_0}$.
The new trial state \eqref{Ansatz_SEMFExc_n} can thus be expressed as a superposition of symmetry-projected Slater determinants:
\begin{subequations} \label{Ansatz_SEMFExc_nx}
\begin{equation} 
  \ket{\Psi_T^{JM\nu}} = \sum_{\a=0}^\nu x^\nu_\a \ket{\Psi^{JM\a}_0},
\end{equation}
with
\begin{equation}
  x^\nu_\a = -\sum_{\b=0}^{\nu-1} (F^{-1})_{\a\b} F_{\b\nu}, 
\end{equation}
\end{subequations}
for $\a=0\to \nu-1$, and $x^\nu_\nu = 1$.

Finally, the phaseless QMC scheme offers an approximation $\ket{\tPG^{JM\nu}}$ for the $\nu$-th excited state of spin $J$ in the vicinity of $\ket{\Psi_T^{J\nu}}$ with energy deduced from Eqs. \eqref{Ph_scheme} and \eqref{Ex_nonyrast_Eg}:
\begin{equation} \label{MixE_SMPPh2}
  \tEG^{J\nu} \eqtauinf \dfrac{\moyE\Bigl[\tP_\t\Re\qmoy{\H}_{\Psi_{T}^{J\nu}\Phi_\t^\nu}\Bigr]}{\moyE\Bigl[\tP_\t\Bigr]}.
\end{equation}
The biased weight $\tP_\t$ of each realization evolves according to Eq. \eqref{Ph_scheme_Pi} with $\tP_0= \ov{\Psi_T^{J\nu}}{\Phi^\nu_0}$, and the orbitals of the walkers $\ket{\Phi_\t^\nu}$ undergo a Brownian motion governed by the one-body representation of the stochastic propagator \eqref{ev_disc_orb} with
\begin{equation} \label{Trial_stateExc}
  \ket{\Psi_T^{J\nu}} = \sum_{\a=0}^\nu x_\a^\nu \sum_{K,K'} C^{J\nu*}_K C^{J\a}_{K'} \op{P}^J_{KK'} \ket{\Phi^\a_0},
\end{equation}
the wave function arising from Eq. \eqref{Ansatz_SEMFExc_nx} and the well-known properties $(\op{P}^J_{MK})^\dagger = \op{P}^J_{KM}$ and $\op{P}^J_{MK}\op{P}^{J'}_{K'M'}$ $ = \delta_{JJ'}\delta_{KK'}\op{P}^J_{MM'}$ of the projection operators.
%
%
%=======================================================================================================
\subsection{Physical observables}
%=======================================================================================================
%
The physical observables of interest in shell model studies are irreducible tensor operators $\A_{kq}$ of rank $k$.
In a state characterized by angular momentum quantum numbers $J$, $M$, solely the $q=0$ components have a non-zero expectation values.
With our QMC approach, none is ensured to be exactly reproduced --- except, obviously, the squared spin and its third component --- because of the bias introduced by the phaseless approximation.
At best, one can achieve the expectation value $\qmoy{\A_{k0}}_{\tPG^{JM\nu}}$ in the state $\ket{\tPG^{JM\nu}}$ by means of two independent populations of walkers to sample the ket and the bra, but such a strategy entails large statistical fluctuations.
Hence, QMC calculations usually rely on the well-known mixed estimate \cite{I_GFQMC1} that provides an approximate expectation value,
\begin{equation} \label{mixed_est}
  \qmoy{\A_{k0}}_{\mathrm{mix}} = \dfrac{\Re\brkt{\Psi_T^{JM\nu}}{\A_{k0}}{\tPG^{JM\nu}}}{\Re\ov{\Psi_T^{JM\nu}}{\tPG^{JM\nu}}}.
\end{equation}
For operators commuting with $\H$, it is identical to the true expectation.
For other observables, it may be corrected by the extrapolate estimate \cite{I_GFQMC1} ($\qmoy{\A}_T$ denotes the expectation value of $\A$ in the trial state)
\begin{equation} \label{Ext_est}
  \qmoy{\A_{k0}}_{\mathrm{ext}} = 2\qmoy{\A_{k0}}_{\mathrm{mix}} -\qmoy{\A_{k0}}_T,
\end{equation}
that is one order of magnitude better in the difference $\ket{\tPG^{JM\nu}}$ $-$ $\ket{\Psi_T^{JM\nu}}$ than Eq. \eqref{mixed_est}.

Nonetheless, a particular attention has to be paid to the QMC reconstruction of the mixed estimator in the case of non-yrast states.
Let us consider the \emph{exact} normalized matrix element from which the mixed estimator is derived
\begin{equation} \label{mixed_est_elem}
 \begin{split}
  \qmoy{\A_{k0}} & _{\Psi_T^{JM\nu}\PG^{JM\nu}}
  = \dfrac{\brkt{\Psi_T^{JM\nu}}{\A_{k0}}{\PG^{JM\nu}}}{\ov{\Psi_T^{JM\nu}}{\PG^{JM\nu}}}\\
 & \proptauinf \dfrac{\brkt{\Psi_T^{JM\nu}}{\A_{k0}(\hat{1}-\P^{JM\nu}_\ex)\e^{-\t\H}}{\Psi_0^{JM\nu}}}{\brkt{\Psi_T^{JM\nu}}{\e^{-\t\H}}{\Psi_0^{JM\nu}}}.
  \end{split}
\end{equation}
When $\nu>0$ and for the energy (or any scalar observable), the factor $(\hat{1}-\P^{JM\nu}_\ex)$ in the trial wave function \eqref{Ansatz_SEMFExc_n0} guarantees that the imaginary-time propagation of the initial state $\ket{\Psi_0^{JM\nu}}$ converges to $\ket{\PG^{JM\nu}}$.
Otherwise, it does not commute with $\A_{k0}$ generally and therefore has to be reintroduced according to Eq. \eqref{mixed_est_elem}.
As before, the rotational invariance and the general form \eqref{Ansatz_SEMF_n} of the projected determinants allows to cast Eq. \eqref{mixed_est_elem} in the form
\begin{multline} \label{mixed_est_elem2}
  \qmoy{\A_{k0}}_{\Psi_T^{JM\nu}\PG^{JM\nu}} \\
  \proptauinf \dfrac{\sum_K C^{J\nu}_K \brkt{\Psi_T^{JM\nu}}{\A_{k0}\op{P}^J_{MK}(\hat{1}-\P^{JK\nu}_\ex)\e^{-\t\H}}{\Phi_0^{\nu}}}{\brkt{\Psi_T^{JM\nu}}{\e^{-\t\H}}{\Phi_0^{\nu}}}.
\end{multline}
In the framework of the phaseless QMC scheme, the trial state becomes \eqref{Ansatz_SEMFExc_n} and the exact projector $\P^{JK\nu}_\ex$ is replaced by $\P^{JK\nu}_T$.
One can then immediately check that
\begin{equation}
  (\hat{1}-\P^{JK\nu}_T)\ket{\Psi_T^{J\nu}} = \ket{\Psi_T^{J\nu}}.
\end{equation}
The walkers being constrained to remain in the vicinity of $\ket{\Psi_T^{J\nu}}$, this implies that the action of $(\hat{1}-\P^{JK\nu}_T)$ approximatively reduces to identity in the stochastic reformulation of the imaginary-time propagation in Eq. \eqref{mixed_est_elem2}, which
leads to
\begin{multline} \label{mixed_est2}
  \qmoy{\A_{k0}}_{\mathrm{mix}} \eqtauinf 
  \dfrac{1}{\moyE\Bigl[\tP_\t\Bigr]}\moyE\Bigl[\tP_\t \Re \Bigl(\sum_\a (x_\a^\nu)^* \times \\
  \sum_{K,K'}  C^{J\a*}_K C^{J\nu}_{K'}\brkt{\Phi^\a_0}{\op{P}^J_{KM}\A_{k0}\op{P}^J_{MK'}}{\Phi_\t^\nu}/\ov{\Psi_T^{J\nu}}{\Phi_\t^\nu} \Bigr)\Bigr],
\end{multline}
for the mixed estimate.

This expression simplifies to a form similar to Eq. \eqref{MixE_SMPPh2} for scalar observables.
For the others, it is no longer possible to permute the operators $\A_{k0}$ and $\op{P}^J_{MK'}$, which means that a double integral over the Euler's angles becomes \emph{a priori} needed.
Still, we may reduce the calculation to a unique integration, tremendously easier to compute.
Indeed, after some algebra involving the behaviour of tensor observables under rotation and the reduction theorem of Wigner's $D$-matrices \cite{III_Messiah}, the product of the three operators in Eq. \eqref{mixed_est2} turns into
\begin{equation} \label{POP_to_Po}
  \op{P}^J_{KM}\A_{k0}\op{P}^J_{MK'} = \mathcal{C}^{JM}_{JMk0} \sum_{q=-k}^k \mathcal{C}^{JK'+q}_{JK'kq}\op{P}^J_{KK'+q}\A_{kq},
\end{equation}
where the symbols $\mathcal{C}^{JM}_{jmj'm'}$ stands for the Clebsch-Gordan coefficients coupling $(j,m)$ and $(j',m')$ in the $(J,M)$ channel \cite{III_Messiah}.
Inserting this result into Eq. \eqref{mixed_est2} then provides the mixed estimator of non-scalar observables
\begin{subequations} \label{Mix_tens}
\begin{equation}
  \qmoy{\A_{k0}}_{\mathrm{mix}} \eqtauinf \dfrac{\moy{\tP_\t \Re\Bigl( \sum_\a \bar{x}_\a^\nu(\t) \qmoy{\A_k^{JM\a\nu}}_{\Phi^\a_0\Phi^\nu_\t}\Bigr)}}{\mathbb{E}\Bigl[\tP_\t\Bigr]} ,
\end{equation}
with the effective operator
\begin{equation} \label{Mix_tens_det}
  \A_k^{JM\a\nu} = \mathcal{C}^{JM}_{JMk0}\!\! \sum_{K,K',q} C^{J\a*}_K C^{J\nu}_{K'} \mathcal{C}^{JK'+q}_{JK'kq}\op{P}^J_{K,K'\!+q}\A_{kq},
\end{equation}
\end{subequations}
and $\bar{x}_\a^\nu(\t) = (x_\a^\nu)^*\ov{\Phi^\a_0}{\Phi^\nu_\t}/\ov{\Psi_T^{J\nu}}{\Phi^\nu_\t}$.
%
%
%=======================================================================================================
\subsection{Variational trial state}
%=======================================================================================================
%
The quality of the trial wave function obviously affects the efficiency of the phaseless QMC method: The better the trial state $\ket{\Psi^{JM\nu}_T}$, the more reduced the artificial bias due to the constraint.
This argument naturally favours the choice for the initial and trial wave functions, Eqs. \eqref{Ansatz_SEMF_n} and \eqref{Ansatz_SEMFExc_n} (or equivalently \eqref{Ansatz_SEMFExc_nx}) respectively, of variational solutions obtained by energy minimization in the subspace of Slater determinants after quantum number projection, i.e. by a variation-after-projection (VAP) approach \cite{PRL}.
This strategy is now also adopted to improve the accuracy of constrained-path QMC simulations for the ground state of the Hubbard model \cite{III_Shi_SPCPMC}, for which it has been observed that trial states preserving the intrinsic symmetries of the Hamiltonian accelerate convergence and significantly reduce the systematic errors \cite{III_Shi_SymQMC}.
For the shell model, such a symmetry-restoration approach matches the so-called VAMPIR (variation after mean-field projection in realistic model spaces) one \cite{III_Schmid_VAMPIR}.
In the case of non-yrast states, we rely on an extension of the VAP approach that consists in sequentially determining variational solutions orthogonal to those previously computed for the lower-energy states, in the spirit of the Excited VAMPIR method \cite{III_Schmid_VAMPIR}.
However, we resort to Slater determinants rather than quasi-particle vacua, and the energy optimisation is carried out differently.

Let us assume we already have VAP approximations of the ground state and the $\nu-1$ lowest excited states of angular-momentum $J, M$.
Then seeking the $\nu$-th excited state, we adopt the ansatz \eqref{Ansatz_SEMFExc_n} with the $2J+1$ amplitudes $\{C_K^{J\nu}\}$ (${-J\leq K\leq J}$) and the $A$ single-particle states $\{\ket{\phi_{0,n}^\nu}\}$ (${n=0\to A}$) of $\ket{\Phi^\nu_0}$ as variational parameters.
In addition, to take advantage of the particular decomposition \eqref{QuadFormH} used, we restrict $\ket{\Phi^\a_0}$, $\forall \a$, to Slater determinants factorized into products of general independent-neutron and -proton wave functions, both breaking the symmetries of the Hamiltonian.
Moreover, as long as the considered valence space contains a single major shell, the parity does not require to be restored.
$\ket{\Psi_T^{JM\nu}}$ finally has a good angular momentum, isospin projection, and parity, so that no further restoration is necessary, except eventually for $N=Z$ nuclei.
Note that Eq. \eqref{Ansatz_SEMFExc_n} holds for $\nu\geq 1$ and translates into Eq. \eqref{Ansatz_SEMF_n} for yrast states.

Variation of the energy
\begin{equation} \label{E_excSEMF}
     E^{J\nu}_T =\dfrac{\sum_{\a,\b} x^{\nu*}_\a x^\nu_\b\sum_{K,K'} C^{J\a*}_K C^{J\b}_{K'} \brkt{\Phi^\a_0}{\H\op{P}^J_{KK'}}{\Phi^\b_0}} {\sum_{\a,\b} x^{\nu*}_\a x^\nu_\b\sum_{K,K'} C^{J\a*}_K C^{J\b}_{K'}\brkt{\Phi^\a_0}{\hat{P}^J_{KK'}}{\Phi^\b_0}}.
\end{equation}
with respect to $C^{J\nu*}_K $ directly yields the following generalized eigenvalue problem
\begin{multline}
  \sum_{K'} C^{J\nu}_{K'} \brkt{\Phi^\nu_0}{\op{P}^J_{KM}\Qnu_T\H\Qnu_T\op{P}^J_{MK'}}{\Phi^\nu_0} = \\
  E^{J\nu}_T \sum_{K'} C^{J\nu}_{K'} \brkt{\Phi^\nu_0}{\op{P}^J_{KM}\Qnu_T\op{P}^J_{MK'}}{\Phi^\nu_0}
\end{multline}
in which $\Qnu_T=(\hat{1}-\P^{JM\nu}_T)$ (for conciseness, we omit the angular-momentum quantum numbers in the new notations).
Developing the matrix elements in both sides provides a form of this equation that does not involve the quantum number $M$ and that is suitable for numerical implementation.

On the other hand, energy minimisation with respect to $\ket{\Phi^\nu_0}$ does no longer lead to a Hartree-Fock-like equation, unlike the case of yrast states \cite{PRL}.
Indeed, as the ansatz \eqref{Ansatz_SEMFExc_n} represents a superposition of independent Slater determinants, the expectation value of the Hamiltonian does not depend any more solely on the one-body density matrix associated with $\ket{\Phi^\nu_0}$.
This difficulty is avoided with the help of the Thouless parametrization \cite{III_Blaizot_Ripka}
\begin{equation}
  \ket{\Phi^\nu_0} = \mathcal{M} \exp\Bigl( \sum_{\bar{n}n} Z_{\bar{n}n} \cc{\bar{n}}\ca{n}\Bigr) \ket{\Phi^\mathrm{r}},
\end{equation}
$\cc{\bar{n}}$ ($\ca{n}$) being the creation (destruction) operator of an unoccupied (occupied) single-particle state $\ket{\phi_{\bar{n}}^{\mathrm{r}}}$ ($\ket{\phi_n^{\mathrm{r}}}$) of a fixed Slater determinant $\ket{\Phi^\mathrm{r}}$ non-orthogonal to $\ket{\Phi_0^\nu}$.
In this way, the projected energy $E^{J\nu}_T$ becomes a complex function of the particle-hole amplitudes $Z_{\bar{n}n}$, and reaches a minimum value when all the components $\partial E^{J\nu}_T/\partial Z_{\bar{n}n}^*$ of its gradient vanish identically.
We now detail their calculation through the introduction of the projected-energy and -overlap matrices \cite{note}
\begin{gather}
  \mathcal{H}_{\a\b} =\brkt{\Psi^{JM\a}_0}{\H}{\Psi^{JM\b}_0} =\int\!\!\d\Om\, X^{\a\b}_\Om \mathcal{O}^{\a\b}_\Om \mathcal{E}^{\a\b}_\Om, \\
  \mathcal{N}_{\a\b} = \ov{\Psi^{JM\a}_0}{\Psi^{JM\b}_0} =\int\!\!\d\Om \, X^{\a\b}_\Om \mathcal{O}^{\a\b}_\Om .
\end{gather}
They allow to cast the variational energy in the form
\begin{equation} \label{E_excSEMF2}
   E^{J\nu}_T = \dfrac{\sum_{\a,\b} x^{\nu*}_\a x^\nu_\b \mathcal{H}_{\a\b}}{\sum_{\a,\b} x^{\nu*}_\a x^\nu_\b \mathcal{N}_{\a\b}}.
\end{equation}
The coefficient
\begin{equation}
   X^{\a\b}_\Om= \sum_{K,K'} C^{J\a*}_K C^{J\b}_{K'} D^{J*}_{KK'}(\Om),
\end{equation}
contains all the dependence on angular momentum.
The energy kernel between the Slater determinants $\ket{\Phi^\a_0}$ and $\op{U}_\Om\ket{\Phi^\b_0}$, i.e.
\begin{equation}
  \mathcal{E}^{\a\b}_\Om = \dfrac{\brkt{\Phi^\a_0}{\H\op{U}_\Om}{\Phi^\b_0}}{\brkt{\Phi^\a_0}{\op{U}_\Om}{\Phi^\b_0}}, 
\end{equation}
may be evaluated through the extended Wick's theorem \cite{III_Blaizot_Ripka}.
It corresponds to the usual Hartree-Fock energy functional but in terms of  the transition one-body density matrix
\begin{equation} \label{trans_density_mat}
  \R_\Om^{\a\b} = \sum_{n,p} U_\Om\ket{\phi^\b_{0,n}}([f^{\a\b}_\Om]^{-1})_{np}\bra{\phi^\a_{0,p}}.
\end{equation}
$(f^{\a\b}_\Om)_{np}= \brkt{\phi^\a_{0,n}}{U_\Om}{\phi^\b_{0,p}}$ refers to the matrix of overlaps between the orbitals.
Moreover,
\begin{equation}
  \mathcal{O}^{\a\b}_\Om = \brkt{\Phi^\a_0}{\op{U}_\Om}{\Phi^\b_0} ,
\end{equation}
is given by the determinant of this matrix \cite{III_Blaizot_Ripka}.
The derivative of $\mathcal{H}_{\a\b}$ with respect to the Thouless amplitudes is greatly simplified by noting that the single-particle effective Hamiltonian $h[\R_\Om^{\a\b}]$ is recovered for the gradient of the Hartree-Fock energy $\mathcal{E}^{\a\b}_\Om$ according to the elements of $\R_\Om^{\a\b}$.
Consequently, one is left with
\begin{multline}
  \dfrac{\partial\mathcal{H}_{\a\b}}{\partial Z_{\bar{n}n}^*} = \delta_{\a\nu} \!\int\!\!\d\Om\,X^{\a\b}_\Om \mathcal{O}^{\a\b}_\Om \bra{\phi_{\bar{n}}^\mathrm{r}}\Bigl[\mathcal{E}^{\a\b}_\Om\R^{\a\b}_\Om \Bigr. \\
  + \Bigl. (1-\R_\Om^{\a\b})h[\R_\Om^{\a\b}]\R_\Om^{\a\b}\Bigr] \ket{\phi_n^\mathrm{r}}, 
\end{multline}
and similarly
\begin{equation}
  \dfrac{\partial\mathcal{N}_{\a\b}}{\partial Z_{\bar{n}n}^*} = \delta_{\a\nu}\!\int\!\!\d\Om\, X^{\a\b}_\Om \mathcal{O}^{\a\b}_\Om \brkt{\phi_{\bar{n}}^\mathrm{r}}{\R_\Om^{\a\b}}{\phi_n^\mathrm{r}}.
\end{equation}
Then, defining the $\nu$-dimensional vector $y^\nu$ of components
\begin{equation}
  y^\nu_\a = -\sum_{\b=0}^{\nu-1} (F^{-1})_{\a\b} (\mathcal{H}x^\nu)_\b,
\end{equation}
($\a=0\to \nu-1$), the gradient may finally be written as:
\begin{subequations}
\begin{multline}
 \dfrac{\partial E^{J\nu}_T}{\partial Z_{\bar{n}n}^*} = \sum_{\a=0}^\nu x_\a^\nu \int\!\!\d\Om X^{\nu\a}_\Om \mathcal{O}^{\nu\a}_\Om \brkt{\phi_{\bar{n}}^{\mathrm{r}}}{\mathcal{T}^{\nu\a}_\Om}{\phi_n^{\mathrm{r}}} \\
  + \sum_{\a=0}^{\nu-1} y_\a^\nu \int\!\!\d\Om X^{\nu\a}_\Om \mathcal{O}^{\nu\a}_\Om \brkt{\phi_{\bar{n}}^{\mathrm{r}}}{\R_\Om^{\nu\a}}{\phi_n^{\mathrm{r}}},
\end{multline}
where we have employed the shorthand notation
\begin{equation}
  \mathcal{T}^{\a\b}_\Om = (1-\R_\Om^{\a\b})h[\R_\Om^{\a\b}]\R_\Om^{\a\b} \\ + (\mathcal{E}_\Om^{\a\b} - E^{J\nu}_T)\R_\Om^{\a\b} .
\end{equation}
\end{subequations}
%
%
%
%
%=======================================================================================================
\section{Proof-of-principle results}\label{Sect_4}
\begin{figure}[!t]
\begin{center}
\includegraphics[scale=0.34,angle=-90]{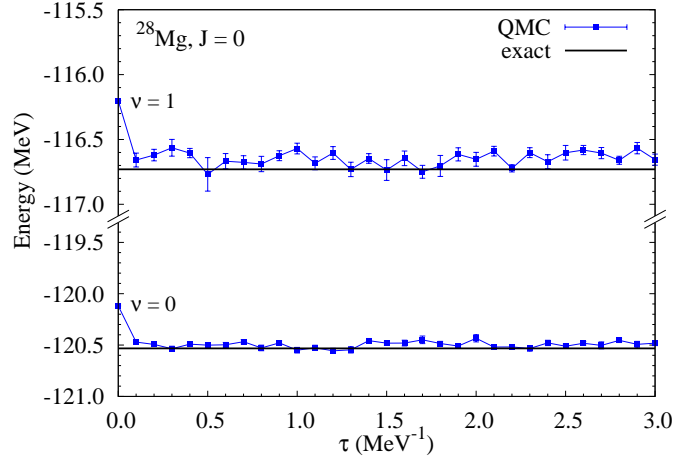}
\caption{\label{Conv}\textbf{(Color online)} Imaginary-time evolution of phaseless QMC energies (blue dots) of the two first $J=0$ states of ${}^{28}$Mg. Exact values are also shown (black lines).}
\end{center}
\end{figure}

%=======================================================================================================
%
This section reports the first applications of the phaseless QMC approach for the shell model extended to non-yrast states.
We address here systems for which the exact diagonalization of the Hamiltonian is possible as benchmarks.
As first example, we focus on the two lowest $J=0$ states of ${}^{28}$Mg whose energies as obtained in the $sd$ space with the USD effective interaction \cite{III_USD} are plotted in Figure \ref{Conv} in function of the imaginary time.
We notice that the phase problem is well managed since there is no explosion of error bars during the dynamics.
Furthermore, the convergence is reached very quickly at $\t \approx 0.3$ MeV${}^{-1}$, that is with only 30 steps (details about the simulations are given in appendix).
But the most important observation is that the signal remains stable a long time after, which shows that the excited state does not collapse to the yrast state as one could expect through a sufficiently long \emph{exact} propagation.
This key future of the method ensues from the phaseless constraint \eqref{Ph_scheme_Pi}.
Indeed, it forces the populations of walkers describing each state to have collectively a strictly positive overlap with their respective trial wave functions that are, in our implementation, orthogonal.

Let us now consider the same three $sd$-shell nuclei than in Ref. \cite{PRL}: the $N=Z$ odd-odd nucleus ${}^{26}$Al, the odd-mass nucleus ${}^{27}$Na, and the even-even nucleus ${}^{28}$Mg.
\begin{figure}[!t]
\begin{center}
\subfigure{\includegraphics[scale=0.4,angle=-90]{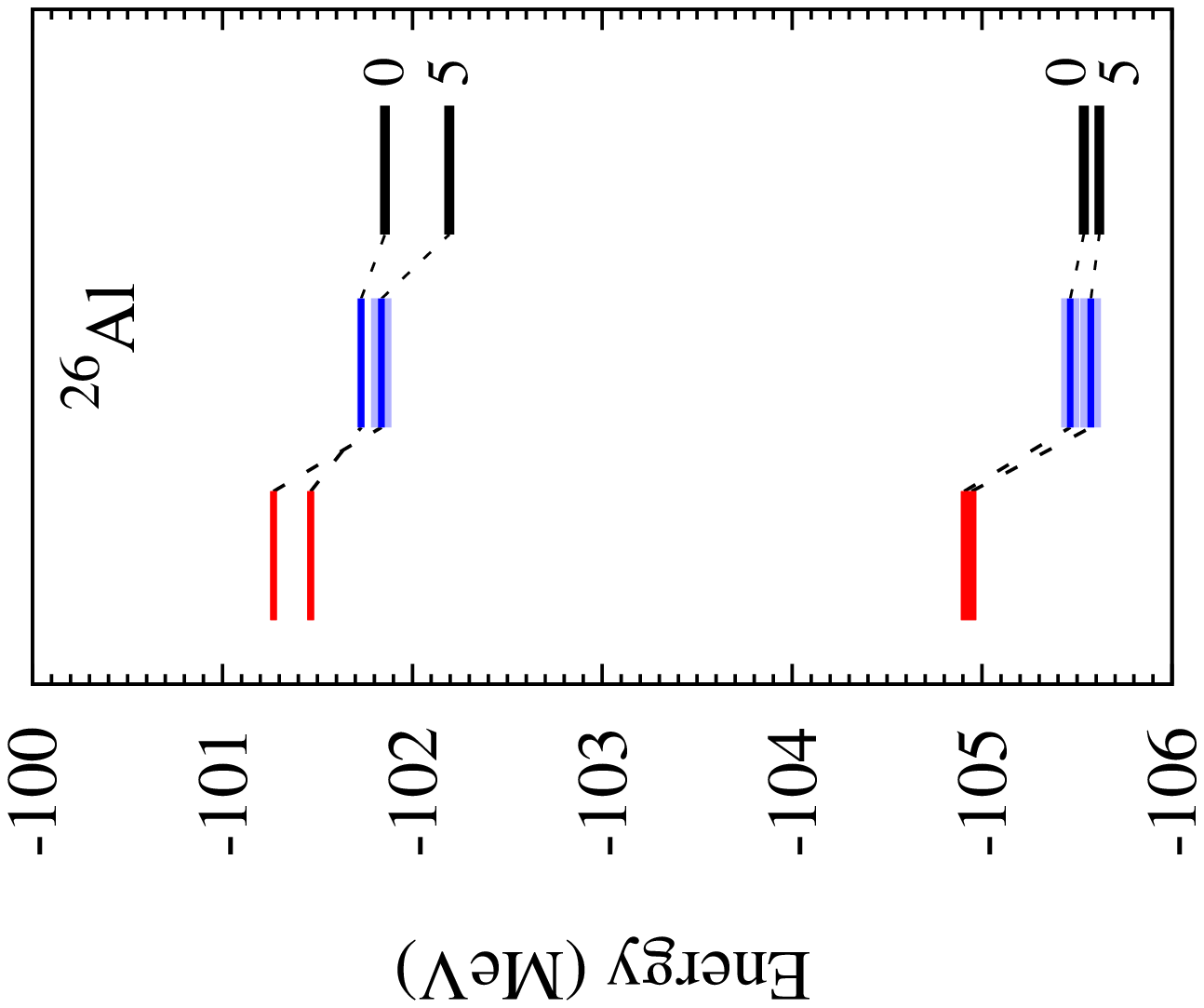}}\\
\subfigure{\includegraphics[scale=0.4,angle=-90]{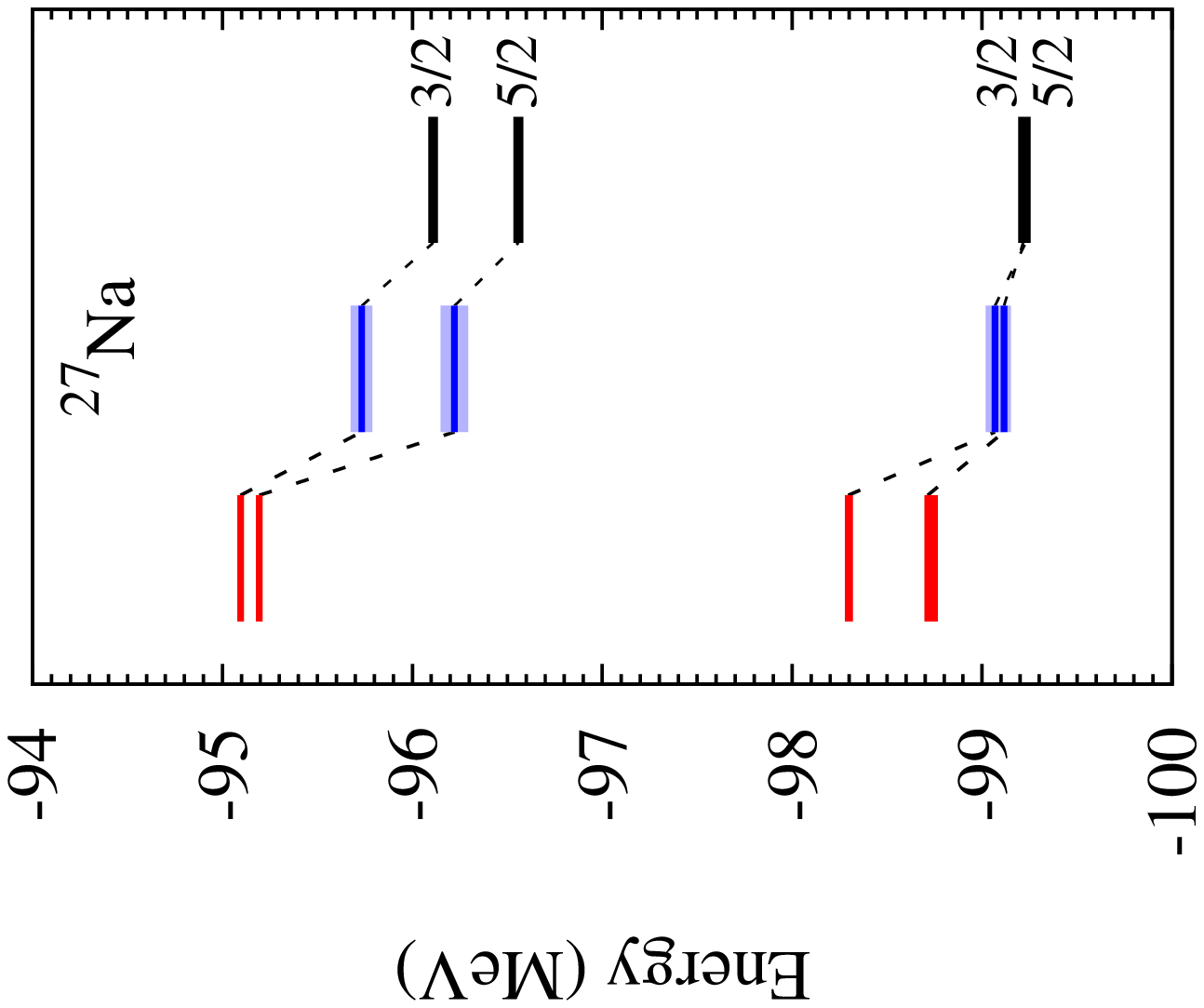}}\\
\subfigure{\includegraphics[scale=0.4,angle=-90]{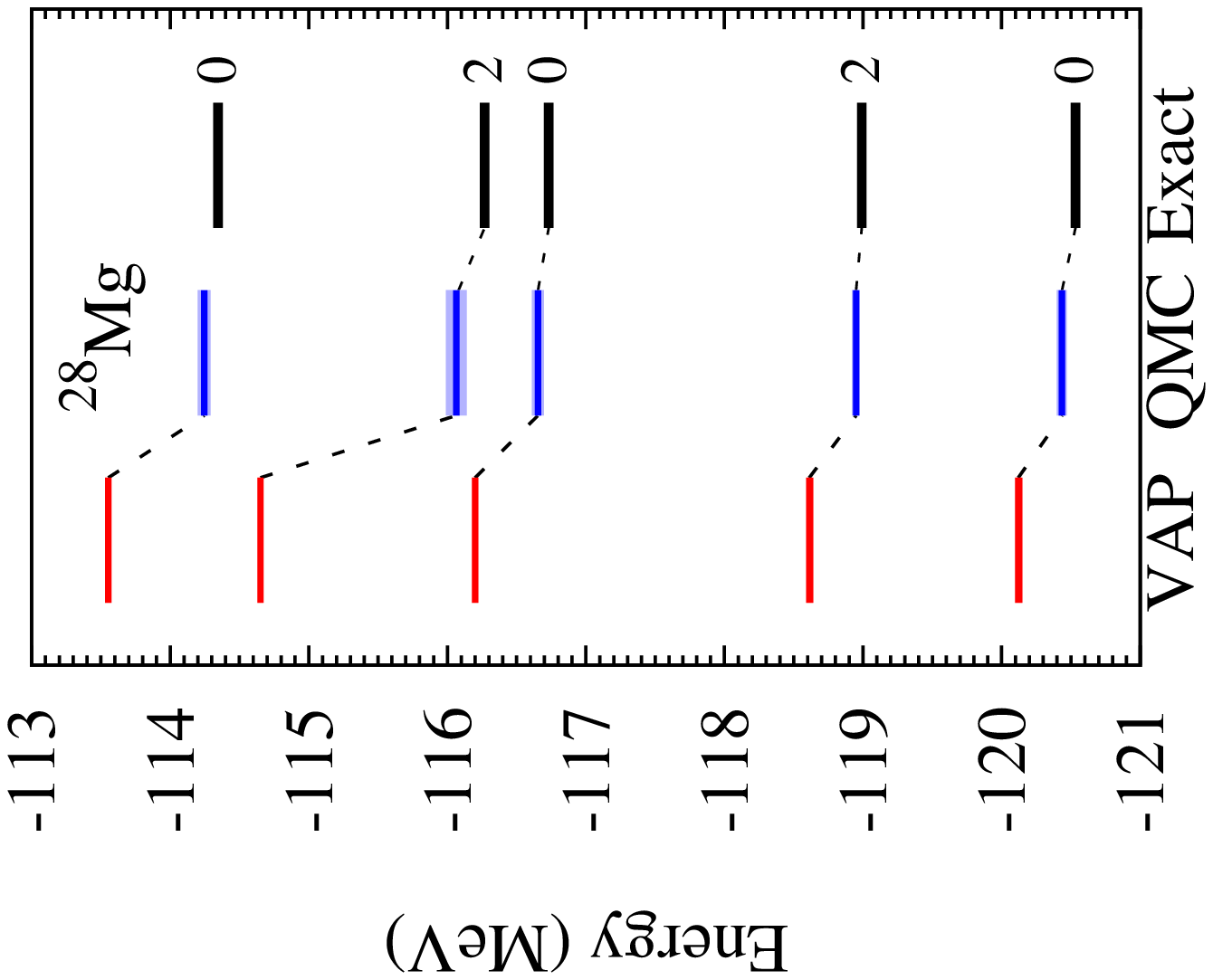}}
\caption{\label{EgExcQMC}\textbf{(Color online)} Binding energies of the two lowest yrast states and of the associated first excited states for the three considered $sd$-shell nuclei, as obtained with the (extended) VAP and phaseless QMC approaches and compared to exact values. The lighter areas around the QMC results represent the statistical errors.}
\end{center}
\end{figure}
Figure \ref{EgExcQMC} displays the binding energies of excited states of same spins than the two lowest yrast states, calculated \textit{via} the extended VAP method and the extended phaseless QMC scheme.
Also shown are the associated yrast energies.
Note that the isovector states in ${}^{26}$Al ($J=0$) have been recovered by way of calculations for the isobaric analogue nucleus ${}^{26}$Mg.
First, we remark that although the variational principle does not apply to the non-yrast cases, we have always obtained VAP energies greater than the exact ones.
We observe that the VAP results for the non-yrast energies approximate rather well the exact values, but in some cases not as well as for the yrast states (see for example the second $J=2$ state in ${}^{28}$Mg whose VAP energy differs from the exact one by 1.6 MeV whereas this discrepancy is around 1 MeV for the yrast levels), which reflects that the VAP wave function then contains more important components onto the higher excited states.
Nonetheless, despite this possible loss of accuracy, the VAP approach still offers a good trial state for guiding and constraining the stochastic paths in the phaseless QMC scheme as the results actually turn out to be of the same quality than in the yrast case:
The phaseless binding energies agree remarkably well with the values from exact diagonalization, the root mean square deviation do not exceed  280 keV with well controlled statistical errors of about 50 keV.
The results for ${}^{27}$Na constitutes a particularly good evidence of the efficiency of the method in controlling the phase problem.
Indeed, it represents the most pathological case for QMC simulations because it combines two sources of phase problems that are an odd number of particles, and a `bad-sign' interaction (see Ref. \cite{I_SMMC_PhP}).

\begin{figure}[!t]
\begin{center}
\subfigure{
\includegraphics[angle=-90,scale=0.165]{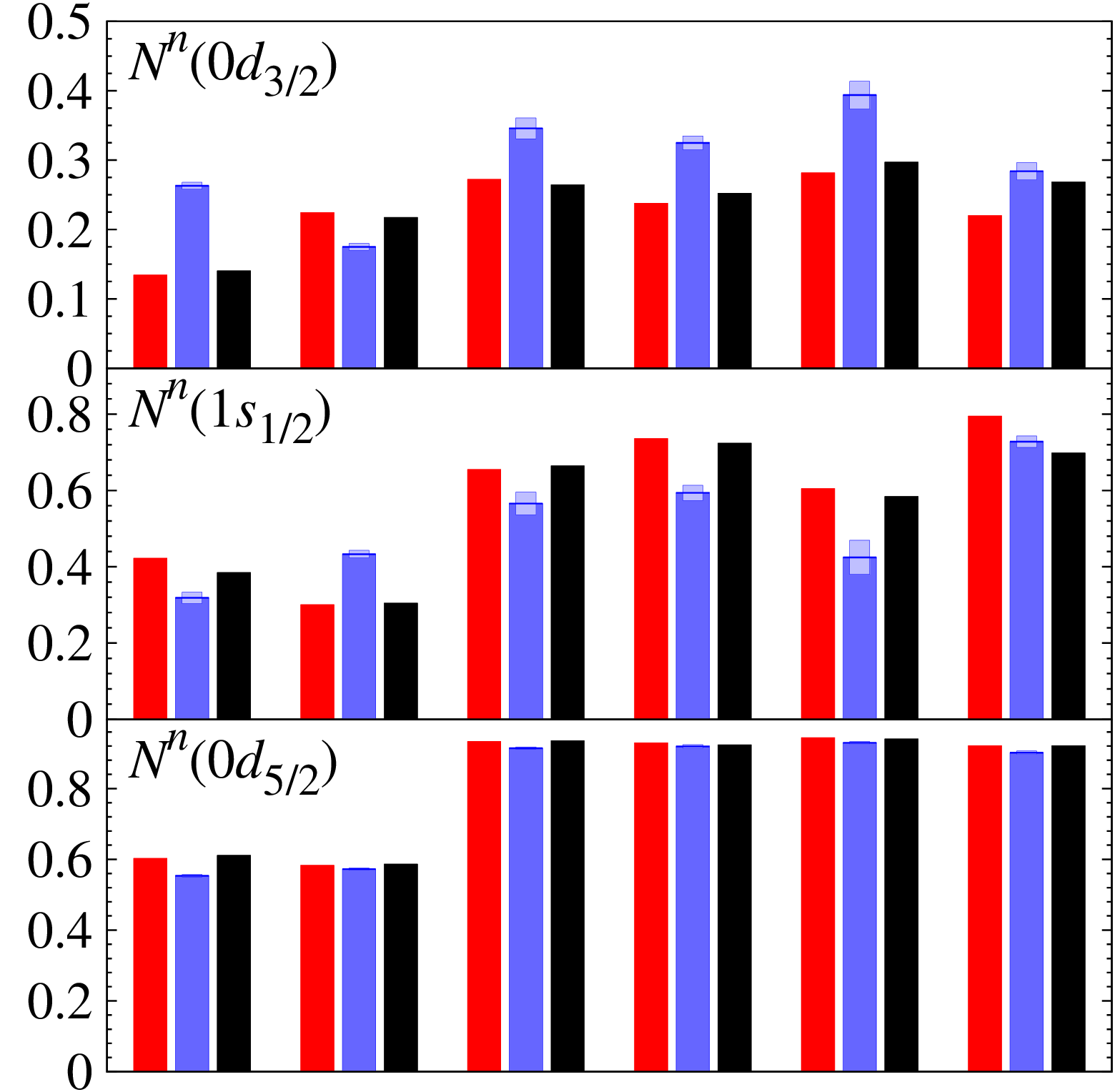}
\includegraphics[angle=-90,scale=0.165]{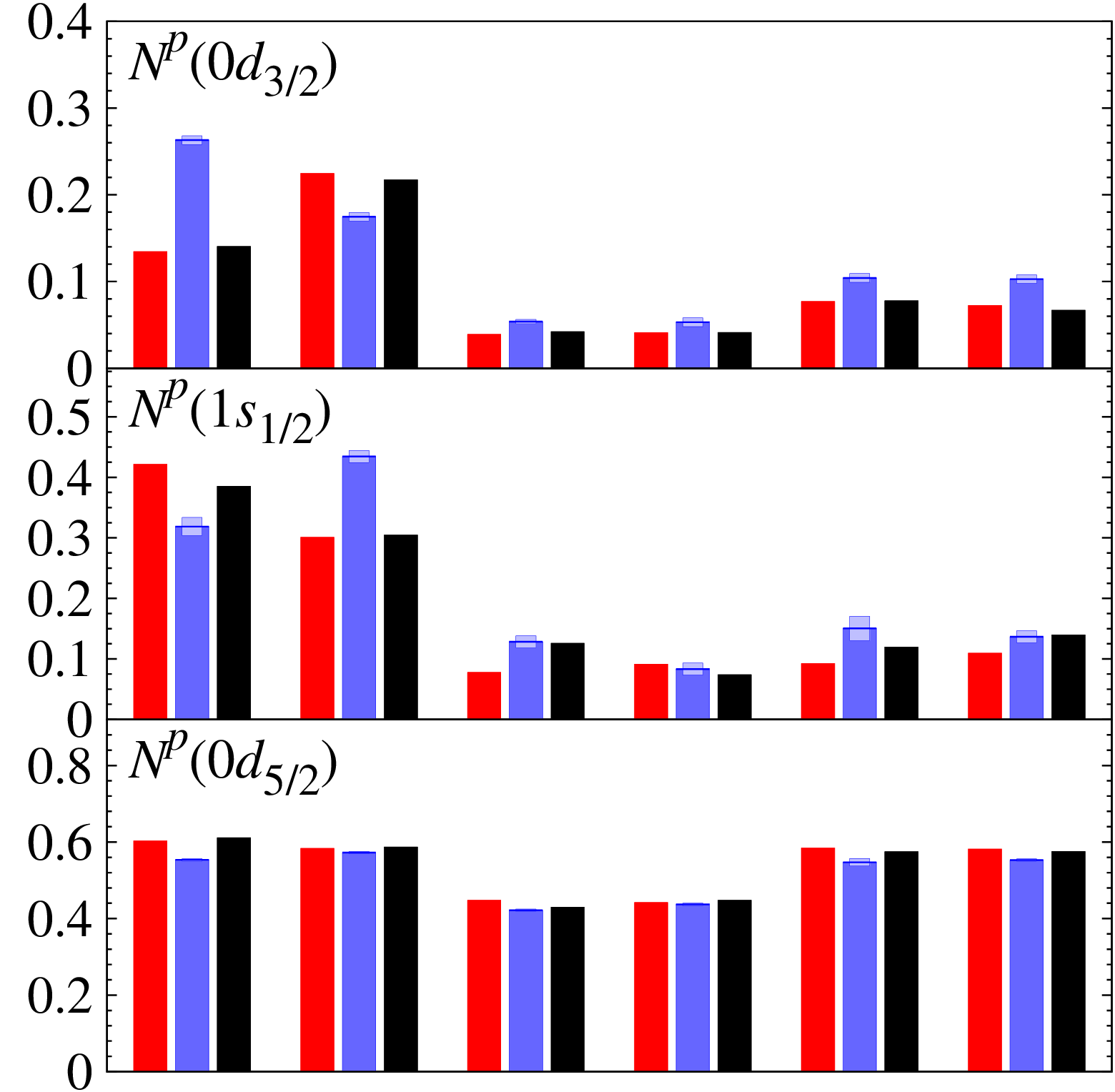}}
\subfigure{
\includegraphics[angle=-90,scale=0.165]{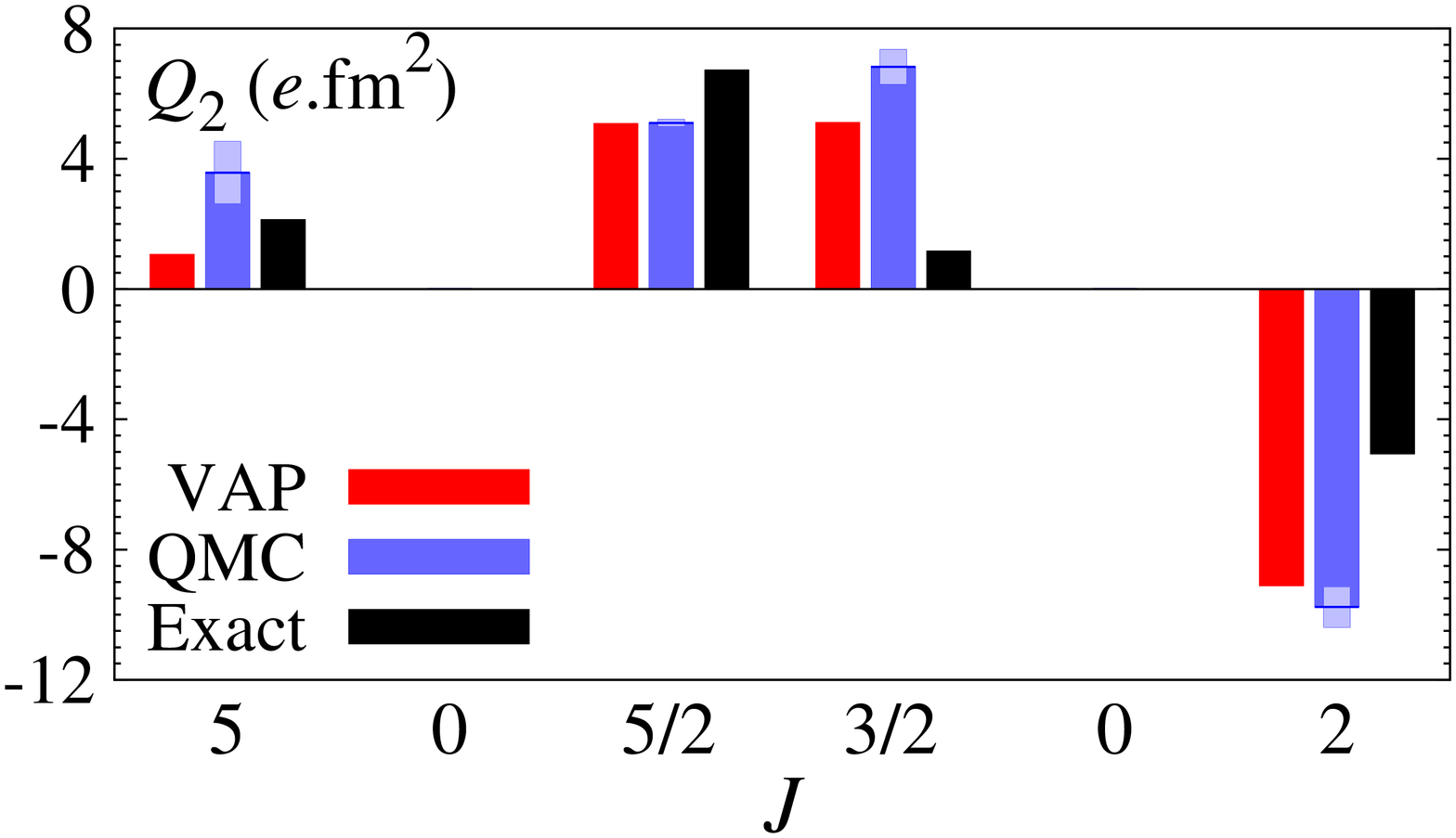}
\includegraphics[angle=-90,scale=0.165]{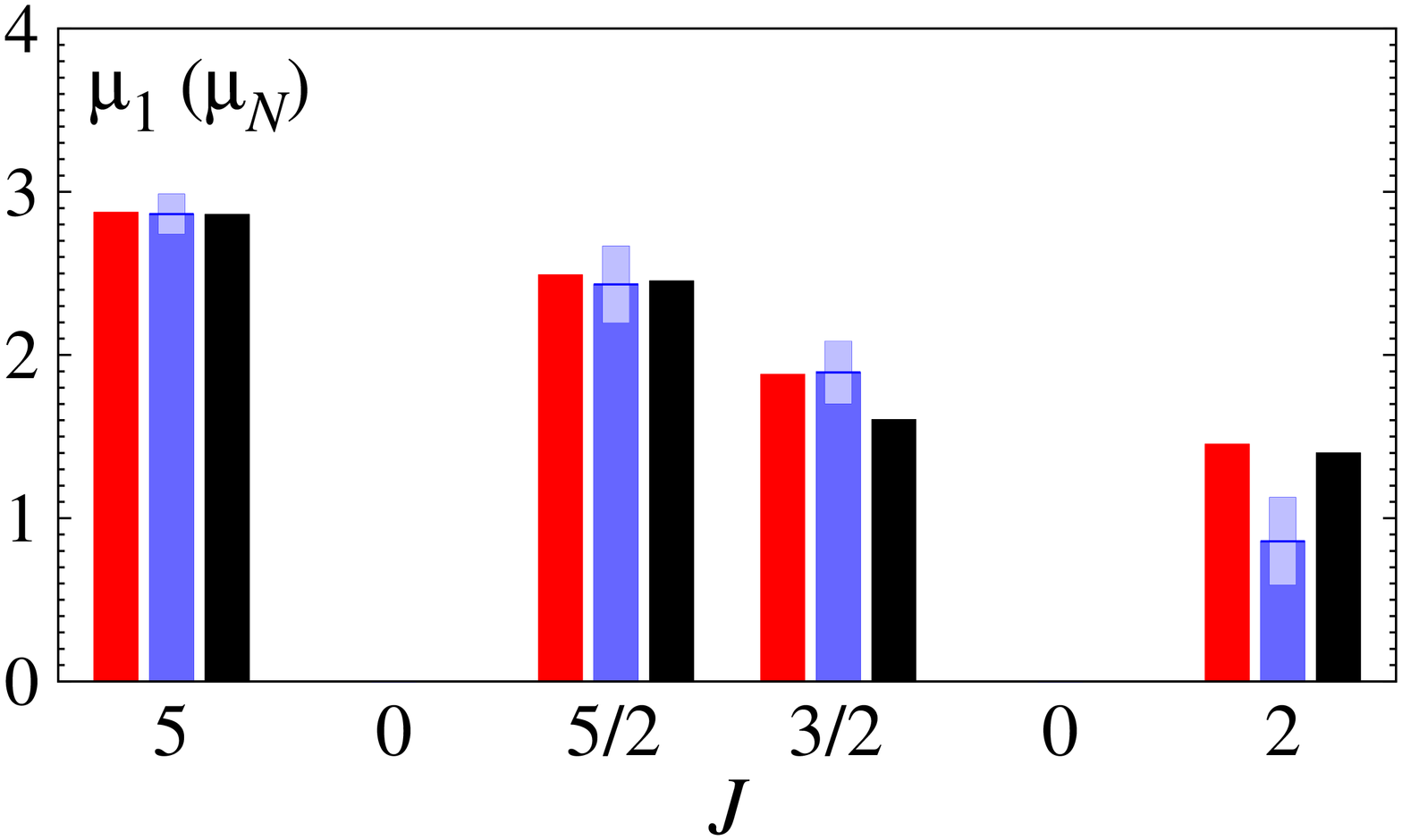}}
\caption{\label{fig_obs}\textbf{(Color online)} VAP, QMC (extrapolate estimate), and exact values of the considered observables for all the first non-yrast states of each spin of the three $sd$-shell nuclei in Fig. \ref{EgExcQMC}: Neutron (upper left) and proton (upper right) normalized occupations, electric quadrupole moment (lower left), and magnetic dipole moment.}
\end{center}
\end{figure}

To further test the ability of the phaseless QMC scheme in reproducing quantities other than energies, we have calculated non-scalar observables that do not commute with the Hamiltonian for which extrapolate estimates \eqref{Ext_est} have to be utilized.
Particularly, we have evaluated the normalized occupations
\begin{equation}
   N^{n(p)} (nlj) = \dfrac{1}{2j+1}\sum_{m=-j}^j \qmoy{\cc{nljm}\ca{nljm}},
\end{equation}
of the orbits $nlj = 0d_{5/2}$, $1s_{1/2}$, $0d_{3/2}$, $\cc{nljm}$ and $\ca{nljm}$ denoting the associated creation and annihilation operators for the neutrons (protons).
The electric quadrupole $Q_2$ and magnetic dipole $\mu_1$ moments are also addressed.
They are respectively defined as the expectation values in an $M=J$ state of
\begin{equation}
\begin{split}
   \hat{Q}_2   &= \sqrt{\dfrac{16\pi}{4}} \sum_i e_i^{\mathrm{eff}} \hat{r}_i^2 Y_{20}( \vec{r}_i), \\
   \hat{\mu}_1 &= \sum_i ( g_i^l \hat{l}_{z,i} + g_i^s \hat{s}_{z,i} ) \mu_N ,
\end{split}
\end{equation}
where $\vec{r}_i$, $\hat{l}_{z,i}$, and $\hat{s}_{z,i}$ stand for the position, momentum, and intrinsic spin of the valence nucleon $i$, and $Y_{20}$ for a spherical harmonic.
The effective charges $e_i^{\mathrm{eff}}$, orbital $g_i^l$ and spin $ g_i^s$ factors are taken from \cite{III_effchargeGfact}.
The results of these observables as computed within the VAP and QMC methods are compiled in Fig. \ref{fig_obs} for the two first non-yrast states of Fig. \ref{EgExcQMC}.
First, we observe that the VAP approach provides a roughly correct agreement with the exact values for any angular momentum.
However, contrary to the binding energies, no significant improvement is found at the outcome of the imaginary-time propagation.
These results probably point out the necessity to go further than extrapolate estimates to evaluate the expectation values of operators that do not commute with the Hamiltonian.

Finally, let us emphasize that for all the considered states we have always recovered exact expectation values of the squared isospin (up to 10${}^{-7}$), for yrast as well as non-yrast states.
This represents a very useful feature of the phaseless QMC approach, especially for $N=Z$ nuclei, since the Wigner-Eckart theorem in isospin space thus brings the possibility to extract the expectation values of observables --- not only the energy --- for isovector states from calculations for the isobaric analogue nucleus.
Such a strategy also holds for the VAP method that produces wave functions almost pure in isospin (relative error on the squared isospin of around 10${}^{-4}$).

%
%
%
%
%===================================================================================================================================
%===================================================================================================================================
\section{Summary and conclusions}\label{Sect_5}
%===================================================================================================================================
%===================================================================================================================================
%
In summary, the phaseless QMC method aims at providing a genuine alternative to the diagonalization of the Hamiltonian matrix, with the purpose to extend the applicability of the nuclear shell model.
The formalism relies on a trial wave function whose roles are (i) to initiate and guide the underlying Brownian motion in order to improve the efficiency of the sampling (ii) to constrain the stochastic paths in order to control the phase problem thanks to the phaseless approximation.
In our application of the approach to the shell model we consider a symmetry-restored trial wave function that absorbs correlations beyond the mean-field level \textit{via} projection before variation.

In this work, we have introduced an extension of the phaseless QMC scheme to reach non-yrast states.
The obtained results show the ability of the method to yield a nearly exact spectroscopy of nuclei, irrespective of the neutron and proton numbers or the interaction.
Nevertheless, further developments are needed to go beyond the standard extrapolate estimate so as to improve the reproduction of observables that do not commute with the Hamiltonian.

Future developments also includes the implementation of electromagnetic transitions and $\beta$-decay probabilities by means of a specific mixed estimator suggested in the context of GFMC techniques \cite{C_mixBE}.
Looking further ahead, a direct consideration of the pairing correlations within the Brownian motion through the propagation of quasi-particle vacua instead of Slater determinants as walkers could reduce the calculation time by increasing the speed of convergence with the imaginary time as well as the number of stochastic realizations.

We gratefully thank S. M. Lenzi for a careful reading of the manuscript and useful comments.
%
%
%
%===================================================================================================================================
%===================================================================================================================================
\appendix
\section*{Appendix: Numerical issues}
%===================================================================================================================================
%===================================================================================================================================
%
This appendix provides some details about the numerical implementation of the phaseless QMC method for the shell model.

The calculations are carried out by considering several independent populations composed of a fixed number $N_w$ of walkers.
They are sampled according to their constrained weight $\tP$ (Eq. \eqref{Ph_scheme_Pi}) via a stochastic reconfiguration algorithm \cite{III_reconif}.
In this way, the mixed estimate \eqref{MixE_SMPPh2} and \eqref{Mix_tens} of an observable is deduced by averaging the mixed estimates reconstructed for each population, and the associated error bars are trivially determined on the basis of the central limit theorem.

The implementation thus contains two tunable parameters that are the imaginary-time step $\Dt$ (see Eq. \eqref{ev_disc_orb}) and the number $N_w$ of walkers.
In order to assess the sensitivity of the results with respect to these quantities, ground-state energies of ${}^{28}$Mg as obtained for different $\Dt$ and $N_w$ are compared in Table \ref{tab_conv}.
The reported values correspond to averages $\bar{E}$ of the energies every 0.1 MeV${}^{-1}$ within the plateau of convergence in Fig. \ref{Conv}.
The upper part shows that, while for $\Dt=0.01$ and $0.025$ MeV${}^{-1}$ the results coincide well and differ from exact diagonalisation by around 35 keV, a systematic deviation emerges for a time step $\Dt=0.1$ MeV${}^{-1}$.
It is explained by the Trotter-Suzuki breakup involved in Eq. \eqref{ev_disc_orb} and whose validity is limited to second order in $\Dt$.
On the lower part, the energies for $N_w=100$ and $N_w=200$ walkers are in agreement up to statistical errors.
Finally, all the calculations presented in this paper have been performed with $\Dt=0.01$ MeV${}^{-1}$ and 30 populations of $N_w=100$ walkers.
\begin{table}[t]
\begin{tabular*}{\linewidth}{@{\extracolsep{\fill}}lccc}
\hline
\hline
\multicolumn{4}{c}{$N_w=100$}\\
\hline
$\Dt$ (MeV${}^{-1}$)& 0.01        & 0.025       & 0.1 \\
$\bar{E}$ (MeV)      & -120.496(9) & -120.504(5) & -120.431(5) \\
\hline
\hline
\end{tabular*}
\begin{tabular*}{\linewidth}{@{\extracolsep{\fill}}lcc}
\multicolumn{3}{c}{$\Dt=0.025$ MeV${}^{-1}$}\\
\hline
$N_w$          & 100         & 200 \\
$\bar{E}$ (MeV) & -120.504(5) & -120.494(5) \\
\hline
\hline
\end{tabular*}
\caption{\label{tab_conv}Average ground-state energies $\bar{E}$ of ${}^{28}$Mg for different imaginary-time steps $\Dt$ and numbers $N_w$ of walkers per population. The exact energy is -120.532 MeV.}
\end{table}   
%
%
%
%===================================================================================================================================
%===================================================================================================================================
% Bibliography
%===================================================================================================================================
%===================================================================================================================================
%
\input{Biblio}
\end{document}

%% file: Biblio.tex
%